\documentclass[10pt, conference, letterpaper]{IEEEtran}

\usepackage{amssymb}
\usepackage{graphicx}
\usepackage{color}
\usepackage{soul} 
\usepackage{mathtools}
\usepackage{url}
\usepackage{caption}
\usepackage{subcaption}
\usepackage{multirow}
\usepackage{xcolor}
\usepackage[colorlinks = true,
            linkcolor = blue,
            urlcolor  = blue,
            citecolor = blue,
            linkbordercolor = blue,
            anchorcolor = blue]{hyperref}

\graphicspath{{./}}

\ifCLASSINFOpdf
\else
\fi

\begin{document}

\title{Experimental Study on Low Power Wide Area Networks (LPWAN) for Mobile Internet of Things}

\author{Dhaval Patel and Myounggyu Won\\
WENS Lab, South Dakota State University, Brookings, SD, United States\\
\{dhaval.patel, myounggyu.won\}@sdstate.edu}%


\maketitle

\begin{abstract}
In the past decade, we have witnessed explosive growth in the number of low-power embedded and Internet-connected devices, reinforcing the new paradigm, Internet of Things (IoT). The low power wide area network (LPWAN), due to its long-range, low-power and low-cost communication capability, is actively considered by academia and industry as the future wireless communication standard for IoT. However, despite the increasing popularity of `mobile IoT', little is known about the suitability of LPWAN for those mobile IoT applications in which nodes have varying degrees of mobility. To fill this knowledge gap, in this paper, we conduct an experimental study to evaluate, analyze, and characterize LPWAN in both indoor and outdoor mobile environments. Our experimental results indicate that the performance of LPWAN is surprisingly susceptible to mobility, even to minor human mobility, and the effect of mobility significantly escalates as the distance to the gateway increases. These results call for development of new mobility-aware LPWAN protocols to support mobile IoT.
\end{abstract}


\IEEEpeerreviewmaketitle

\section{Introduction}
\label{sec:introduction}

Explosive growth in the number of Internet-connected ``things'' in the past decade has driven the emergence of new wireless communication technology: low power wide area network (LPWAN). LPWAN is increasingly gaining popularity from industrial and research communities because of its low-power, long-range, and low-cost communication characteristics. More specifically, it provides long-range communication of up to 10-15 km in rural areas and 2-5 km in urban areas~\cite{centenaro2015long}, and it is highly energy-efficient and inexpensive--the industry is targeting 10+ year battery life~\cite{weightlesslpwan} with a radio chipset cost of less than \$2 and the operating cost of \$1 per device per year~\cite{lpwancost}.

This promising prospect of LPWAN has prompted recent experimental studies on the performance of LPWAN~\cite{petajajarvi2016evaluation}\cite{adelantado2016understanding}\cite{petric2016measurements}. However, IoT devices are getting more mobile as manifested by recent IoT applications (\emph{e.g.,} healthcare~\cite{amendola2014rfid}, automotive sensor networks~\cite{guo2015mobile}, industrial applications~\cite{karnouskos2012soa}, and road conditioning~\cite{casselgren2012model}). IoT devices are increasingly attached and operated in mobile objects like vehicles, trains, and airplanes, \emph{etc.} Furthermore, flexible and wearable sensors are more widely used~\cite{pantelopoulos2010survey}; it is forecasted that there will be more than three billion wearable sensors by 2050~\cite{wearable}.

Many researchers have already stressed the significance of mobile IoT. Stankovic remarked the robustness issue in a mobile environment, \emph{i.e.,} the system stability is impacted by mobility~\cite{stankovic2014research}. Chen \emph{et al.} reported that the IoT services in China are becoming mobile, decentralized, and complex~\cite{chen2014vision}. Mobile IoT for smart cars has been considered~\cite{yau2014intelligent}. Skorin-Kapov \emph{et al.} approached mobile IoT from the perspective of mobile crowdsensing, \emph{i.e.,} collecting data from a large number of mobile sensors~\cite{skorin2014energy}. Mozaffari \emph{et al.} extended the limitation of static IoT by integrating the mobility of UAVs with IoTs~\cite{mozaffari2016mobile}. Rosario \emph{et al.} investigated a routing protocol for mobile IoTs~\cite{rosario2014beaconless}. However, despite the increasing significance and popularity of mobile IoT, little is known about whether LPWAN is a suitable communication standard for those mobile IoT applications.

To fill this knowledge gap, in this paper, we perform a comprehensive real-world experimental study to evaluate, analyze, and characterize the performance of LPWAN in both indoor and outdoor environments with varying degrees of mobility. Consequently, we report three major findings on the performance of LPWAN (in terms of end-to-end delay and packet loss rates): 1) the performance of LPWAN is impacted even by a small degree of mobility (\emph{i.e.,} human mobility); 2) The effect of mobility is greater in an indoor environment; and 3) The longer distance to the gateway further escalates the impact of mobility. These results suggest that new mobility-aware LPWAN protocols need to be developed.



This paper is organized as follows. In Section~\ref{sec:system_design}, we present a brief review of the LPWAN technology and introduce the LPWAN platform that we used for this study. We then present the details of the experimental setup in Section~\ref{sec:experiment_setup}. In Section~\ref{sec:experiment_results}, the experimental results are presented. We then conclude in Section~\ref{sec:conclusion}.

\section{Low Power Wide Area Network: Overview}
\label{sec:system_design}

In this section, we present an overview of the LPWAN technology followed by justifications for selecting a specific LPWAN platform.

Sub-GHz unlicensed ISM bands (\emph{e.g.,} 868 MHz in Europe, and 915 MHz in the U.S.) are used to operate LPWAN. The communication range for LPWAN reaches up to 15 km in rural areas, and up to 5 km in urban areas\cite{centenaro2015long}--some reports the range of up to 30 km in rural areas~\cite{petajajarvi2016evaluation}. This long range of LPWAN is possible with a new physical layer design that allows for significantly high receiver sensitivities, \emph{e.g.,} -130 dBm.

To support the long-range communication of LPWAN, its data rate is necessarily low as a few hundred to thousand bits/sec. Thus, LPWAN is better suited for low-power IoT devices that transmit a small amount of data over a long distance, in contrast to short-range technologies such as Bluetooth and Zigbee. In addition, compared to cellular M2M networks that are designed to cover a large area, LPWAN is more cost-effective due to its low hardware price and no need for subscription for service~\cite{CVZZ16}.

\begin{figure}[!htbp]
\centering
\includegraphics[width=.7\columnwidth]{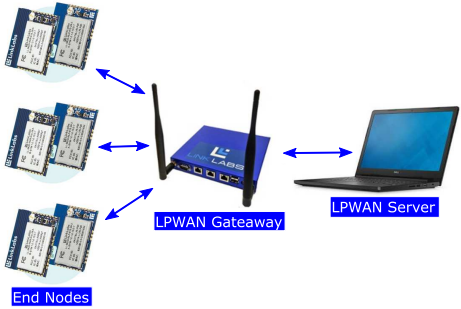}
\caption {Star topology of LPWAN.}
\label{fig:lpwan_system_architecture}
\end{figure}

Most LPWANs are formed based on the star topology where end nodes are directly connected to a gateway that relays data to a LPWAN server (Figure~\ref{fig:lpwan_system_architecture}). This significantly simplifies the network design allowing for high scalability and greater controllability. Currently, there exist a number of commercial platforms for LPWAN, \emph{e.g.,} SIGFOX\textsuperscript{TM}~\cite{Sigfox}, LoRa\textsuperscript{TM}~\cite{LoRa}, Ingenu\textsuperscript{TM}~\cite{Ingenu}, \emph{etc.}

For this experimental study, we adopted the Symphony Link that is built on LoRaWAN which is the LPWAN platform of LoRa Alliance~\cite{Symphony}. Here we present justifications for selecting Symphony Link for this experimental study.

\begin{itemize}
  \item Utilizing per-packet acknowledgement, Symphony Link has lower packet error rates.
  \item Symphony Link is flexible in adjusting the duty cycle allowing us to send more packets at a given time.
  \item It is more flexible in terms of controlling the transmission power and data rates.
  \item The fixed MTU size of 256 bytes of Symphony Link allows us to send packets of varying sizes.
\end{itemize}

\section{Experimental Setup}
\label{sec:experiment_setup}


\begin{figure}[!htbp]
\centering
\includegraphics[width=.98\columnwidth]{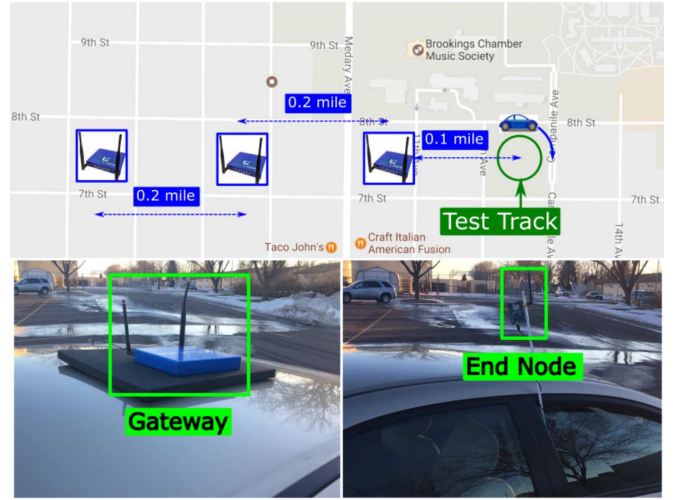}
\caption {Experimental setup.}
\label{fig:experimental_map}
\end{figure}

The experiments were performed in both indoor and outdoor environments. The RX sensitivity of the gateway was -133dBm. The max RX gain of the gateway antenna was -39.4dBm. The height of the gateway and the end node (RXR-27) was around 1.5m in both the indoor and outdoor scenarios. The TX power of the end node between 12dBm and 26dBm was selected based on the adaptive data rate (ADR) scheme of LoRaWAN~\cite{lorawanoverview}. 

The indoor experiments were carried out in the hallway (3m by 130m) on the third floor of a building. A person continuously moved at a normal walking speed from one end to the other of the hallway for 20mins for each measurement (\emph{i.e.,} for each packet size, and for each distance to the gateway). The gateway was initially installed in the middle of the hallway, and it was placed outside the building to increase the distance to the gateway. The distance was varied from 0 to 0.3 miles. To compare the results with the non-mobility case, we fixed the location of the end node and repeated the same experiments.

A vehicle was used as a means to test for high mobility in an outdoor environment. An empty parking lot was exploited for this experiment in which a circular test track was defined to maintain a constant vehicle speed (Figure~\ref{fig:experimental_map}). An end node was installed on top of the vehicle (Figure~\ref{fig:experimental_map}). The vehicle utilizing its cruise control system continuously traveled on the circular test track with varying speed from 5 mph up to 15 mph for 20 mins for each measurement. The gateway was placed at varying distances from the center of the test track, \emph{i.e.,} 0.1, 0.3, and 0.5 miles. Due the low height of the LPWAN gateway and buildings on the campus, the maximum range was about 0.6 and 0.4 miles in the outdoor and indoor settings, respectively.

\section{Experimental Results}
\label{sec:experiment_results}

In this section, we evaluate the performance of LPWAN in mobile environments. The performance was measured in terms of end-to-end delay and packet loss rates. The end-to-end delay refers to the elapsed time from the point where a packet is transmitted until the acknowledgement packet is received from the gateway. The two performance metrics are measured by varying the following parameters: vehicle speed (for outdoor experiments), packet size, and distance to the gateway.

\subsection{Impact of Packet Size - Indoor Mobile Environment}
\label{sec:impact_of_packet_size_indoor}

Results from experiments performed in the indoor mobile environment are presented. For this experiment, human mobility was applied to investigate the impact of mobility with varying packet sizes. The objectives were to understand how the mobility influences the performance of LPWAN and to discover the correlation between the impact of mobility and the packet size. The gateway was placed in the middle of the hallway.

\begin{figure}[!htbp]
\centering
\includegraphics[width=.7\columnwidth]{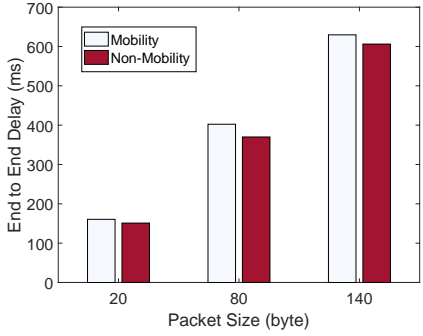}
\caption {Effect of mobility with varying packet sizes (indoor).}
\label{fig:e2e_packet_size_indoor}
\end{figure}

Figure~\ref{fig:e2e_packet_size_indoor} displays the average end-to-end delay for packets that were transmitted for 20 mins. The results indicate that the average end-to-end delay increased as we made the packet size larger. These results are not surprising that coincide with recent research~\cite{adelantado2016understanding}. A key observation was that regardless of the packet size, the average end-to-end delay of the mobility case was consistently greater than the non-mobility scenario. More specifically, compared with the non-mobility case, increases of 5.7\%, 8.9\%, and 3.7\% in the average end-to-end delay were observed for the packet sizes of 20, 80, and 140 bytes, respectively, for the mobility case. Although these differences seem small, interestingly, when the distance to the gateway was greater, the average end-to-end delay for the mobility case significantly increased (Figures~\ref{fig:indoor_distance_1_e2e} and~\ref{fig:indoor_distance_2_e2e}). These results suggest that the end-to-end delay of LPWAN is affected by even minor human mobility, and is more substantially impacted when the distance to the gateway is increased. Another interesting observation was that the packet size did not contribute much to the effect of mobility. It was also worthy to note that the packet loss rates were 0\% regardless of the packet size when the gateway was placed close to the end node (\emph{i.e.,} in the building).

\subsection{Impact of Distance - Indoor Mobile Environment}
\label{sec:impact_of_distance}

As mentioned in Section~\ref{sec:impact_of_packet_size_indoor}, the impact of mobility in an indoor environment becomes more significant when the distance to the gateway increases. This section presents in-depth evaluation of the correlation between the mobility impact and the distance to the gateway. For this experiment, we placed the gateway outside of the building at 0.1 and 0.3 miles away from the building. With the default packet size of 80 bytes, we measured the average end-to-end delay and packet loss rates for both the mobility and non-mobility cases.

\begin{figure}[!htbp]
\centering
\includegraphics[width=.7\columnwidth]{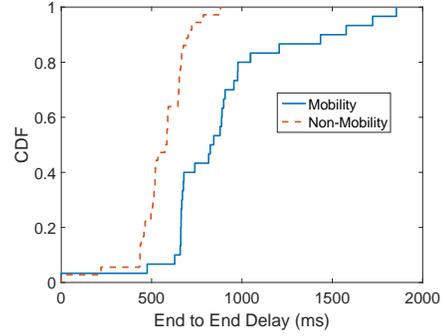}
\caption {Effect of mobility with distance of 0.1 mile (indoor).}
\label{fig:indoor_distance_1_e2e}
\end{figure}

\begin{figure}[!htbp]
\centering
\includegraphics[width=.7\columnwidth]{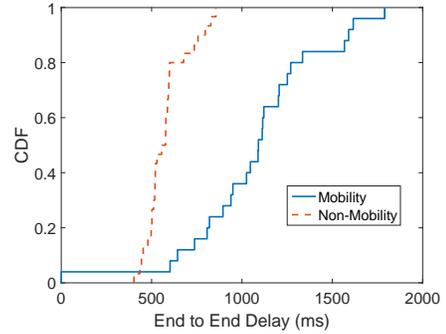}
\caption {Effect of mobility with distance of 0.3 mile (indoor).}
\label{fig:indoor_distance_2_e2e}
\end{figure}

The CDF graphs of the end-to-end delay for the distances of 0.1 mile and 0.3 mile are depicted in Figures~\ref{fig:indoor_distance_1_e2e} and~\ref{fig:indoor_distance_2_e2e}, respectively. When we placed the gateway at 0.1 mile away from the building, the average end-to-end delay significantly increased by 57\%. The average end-to-end delay further increased by 87\% when the distance was increased to 0.3 mile. The two figures display distinctive end-to-end delay differences between the mobility and non-mobility cases.



\begin{figure}[!htbp]
\centering
\includegraphics[width=.7\columnwidth]{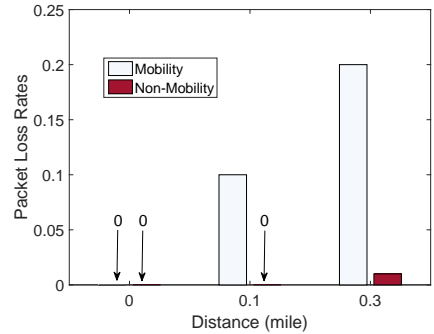}
\caption {Effect of mobility on packet loss rates (indoor).}
\label{fig:indoor_distance_pdr}
\end{figure}

We then measured packet loss rates for varying distances to the gateway. Recall that packet loss rates were 0\% when the gateway was inside the building. The results for the increased distance to the gateway are very interesting (Figure~\ref{fig:indoor_distance_pdr}). As the gateway was placed farther away from the end node, the packet loss rate was significantly impacted even by the minor human mobility. More specifically, for the non-mobility case, the packet loss rate was less than 2\% regardless of the distance to the gateway. However, for the mobility case, the packet loss rates substantially increased, \emph{i.e.,} by up to 10\%, and 20\% for the distances of 0.1 miles, and 0.3 miles, respectively.

\subsection{Impact of Packet Size - Outdoor Mobile Environment}
\label{sec:impact_of_packet_size}

We evaluated the impact of mobility on the performance of LPWAN in the outdoor environment. The gateway was placed 0.1 miles away from the center of the test track. We then measured the average end-to-end delay and packet loss rates by varying the vehicle speed and packet size.

\begin{figure}[!htbp]
\centering
\includegraphics[width=.7\columnwidth]{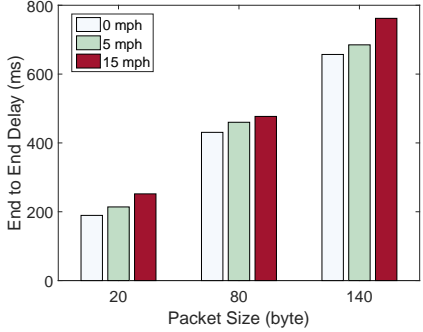}
\caption {Effect of mobility with varying packet sizes (outdoor).}
\label{fig:e2e_speed_packet}
\end{figure}

Figure~\ref{fig:e2e_speed_packet} depicts the results. A strong correlation between the mobility (\emph{i.e.,} vehicle speed) and average end-to-end delay was found: the average end-to-end delay increased as the vehicle speed increased. Similar to the results from the indoor environment, this mobility impact significantly increased as the distance to the gateway increased (Figure~\ref{fig:outdoor_distance_e2e}). Compared with the results from the indoor experiments, we obtained smaller end-to-end delay for the outdoor experiments for all packet sizes due primarily to the signal obstruction.

\begin{figure}[!htbp]
\centering
\includegraphics[width=.7\columnwidth]{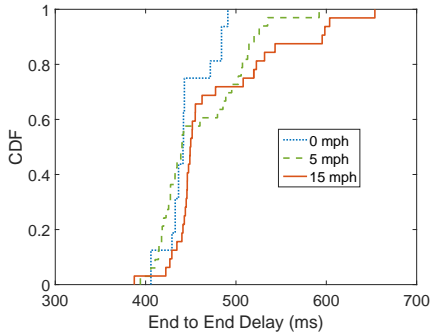}
\caption {The CDF of end-to-end delay for different vehicle speed with packet size of 80 bytes at 0.1 mile (outdoor).}
\label{fig:e2e_speed_packet_cdf}
\end{figure}

The CDF graph of the end-to-end delay in Figure~\ref{fig:e2e_speed_packet_cdf} (for the packet size of 80 bytes) more clearly illustrates the effect of the vehicle speed on the end-to-end delay. As it is shown, at the distance of 0.1 mile, 10\% of the end-to-end delay measurements were greater than 450 ms, 500 ms, and 600 ms for 0 mph, 5 mph, and 15 mph, respectively.

\begin{figure}[!htbp]
\centering
\includegraphics[width=.7\columnwidth]{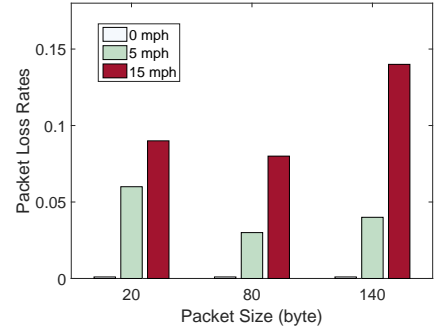}
\caption {Effect of mobility on packet loss rates (outdoor).}
\label{fig:outdoor_pdr_packet_size}
\end{figure}

Figure~\ref{fig:outdoor_pdr_packet_size} displays packet loss rates for different vehicle speeds and packet sizes. It was discovered that there is a strong correlation between the vehicle speed and packet loss rates: the packet loss rates increased as the vehicle speed increased. It was also interesting to note that even the low vehicle speed substantially impacted the packet loss rates. We were not able to find a relationship between the packet size and the packet loss rates in this experiment.

\subsection{Impact of Distance - Outdoor Mobile Environment}
\label{sec:impact_of_transmit_power}

To investigate how the distance to the gateway influences the degree of the mobility impact, we placed the gateway at different distances, \emph{i.e.,} at 0.3 mile and 0.5 mile away from the center of the test track. We then measured the average end-to-end delay and packet loss rates.

\begin{figure}[!htbp]
\centering
\includegraphics[width=.7\columnwidth]{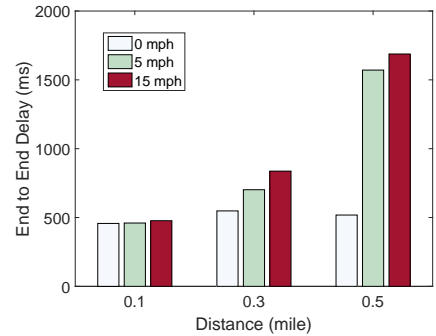}
\caption {Effect of mobility with varying distance (outdoor).}
\label{fig:outdoor_distance_e2e}
\end{figure}

Figure~\ref{fig:outdoor_distance_e2e} depicts the results. As shown, the average end-to-end delay increased with higher vehicle speed regardless of the distance to the gateway. An interesting observation was that the average end-to-end delay more sharply increased with the longer distance to the gateway. More specifically, when the gateway was close to the center of the test track, the effect of mobility was observed but the degree was not substantial in comparison with that for the longer distance to the gateway: when the vehicle speed was increased to 15 mph, increases of up to 4\%, 52\%, and 225\% in the average end-to-end delay were observed for the distances of 0.1, 0.3, and 0.5 miles, respectively.

\begin{figure}[!htbp]
\centering
\includegraphics[width=.7\columnwidth]{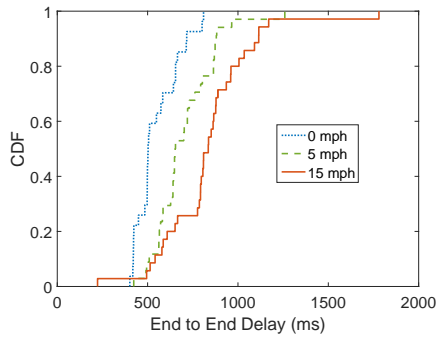}
\caption {The CDF of end-to-end delay for different vehicle speed with packet size of 80 bytes at 0.3 mile (outdoor).}
\label{fig:outdoor_distance_e2e_cdf_dist_3}
\end{figure}

To illustrate the distributions of the end-to-end delay measurements, the CDF graph of the end-to-end delay at the distance of 0.3 mile is displayed in Figure~\ref{fig:outdoor_distance_e2e_cdf_dist_3}. An interesting observation was that, in comparison with the results for the distance of 0.1 mile (Figure~\ref{fig:e2e_speed_packet_cdf}), the gaps between lines are larger indicating more significant increases in the end-to-end delay when the distance to the gateway was longer. We also observed that the mobility impact was greater in an indoor environment compared with the outdoor environment. The reason is that the indoor environment had higher signal disruption due to obstacles and more sources of interference.


\begin{figure}[!htbp]
\centering
\includegraphics[width=.7\columnwidth]{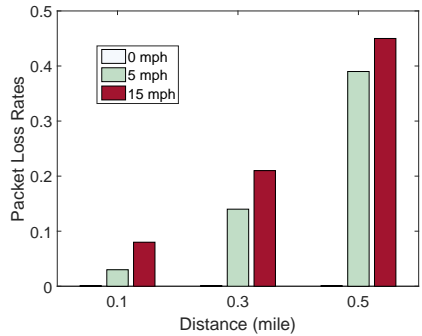}
\caption {Effect of mobility on packet loss rates (outdoor).}
\label{fig:outdoor_distance_pdf}
\end{figure}

Figure~\ref{fig:outdoor_distance_pdf} depicts the results of packet loss rates. There is a clear correlation between the packet loss rates and the vehicle speed in the outdoor environment: as we increased the vehicle speed the packet loss rates increased. It was also observed that this mobility impact became greater when the distance to the gateway increased. Regardless of the distance to the gateway, when the vehicle speed was 0 mph, the packet loss rate was extremely low. Also note that in the outdoor environment, even with the small vehicle speed (\emph{i.e.,} 5 mph), the packet loss rate was significantly affected.



\section{Conclusions}
\label{sec:conclusion}

We have presented a real-world experimental study that revealed the relationship between the mobility and the performance of LPWAN to understand the suitability of LPWAN for mobile IoT. Consequently, we provided rather negative results: LPWAN is easily impacted by mobility, even by minor ones such as human mobility. The impact of mobility dramatically increased depending on the distance to the gateway, the vehicle speed, and whether the end node was placed in an indoor environment. As future work, based on these results, we will develop mobility-aware LPWAN protocols that address this mobility issue.

\bibliographystyle{IEEEtran}
\bibliography{reference}

\end{document}